\documentclass[usegraphicx,usenatbib,useapjfonts,apj]{emulateapj}

\def\fsl{\mbox{$\lambda_{f}$}} 
\def\hubble{\mbox{km sec$^{-1}$ Mpc$^{-1}$}} 
\def\kev{\mbox{keV}} 
\def\kms{\mbox{km/s}} 
\def\kpch{\mbox{$h^{-1}$kpc}} 
 
\def\mpch{\mbox{$h^{-1}$Mpc}} 
\def\msunh{\mbox{$h^{-1}$M$_\odot$}} 
\def\msun{\mbox{M$_\odot$}} 
\def\ome{\mbox{$\Omega_0$}} 
\def\omel{\mbox{$\Omega_\Lambda$}} 
 
\def\sige{\mbox{$\sigma_8$}} 
\def\vmax{\mbox{$V_{\rm max}$}} 
\def\vth{\mbox{$v_{\rm rms}$}} 
\def\mnu{\mbox{$m_{\rm W}$}}  
\def\rvir{\mbox{R$_v$}}
\def\rc{\mbox{r$_{c}$}}
\def\mvir{\mbox{M$_v$}}
\def\cq{\mbox{c$_{1/5}$}}
\def\cv{\mbox{c$_{\rm NFW}$}}
\def\lcdm{\mbox{$\Lambda$CDM}}
\def\funits{\mbox{M$_\odot$pc$^{-3}$/(km/s)$^3$}} 

\def\Anv{\mbox{A256$_{0.0}$}}
\def\Bnv{\mbox{B256$_{0.0}$}}
\def\Cnv{\mbox{C512$_{0.0}$}}
\def\Dnv{\mbox{D512$_{0.0}$}}
\def\Env{\mbox{E512$_{0.0}$}}
\def\Av{\mbox{A256$_{0.1}$}}
\def\Bv{\mbox{B256$_{0.1}$}}
\def\Cv{\mbox{C512$_{0.1}$}}
\def\Dv{\mbox{D512$_{0.1}$}}
\def\Dvb{\mbox{D512$_{0.1}$b}}
\def\Ev{\mbox{E512$_{0.1}$}}

\def\Cvv{\mbox{C512$_{0.3}$}}
\def\Dvvb{\mbox{D512$_{0.3}$b}}
\def\Evv{\mbox{E512$_{0.3}$}}

\def\Dnvll{\mbox{D128$_{0.0}$}}
\def\Dnvl{\mbox{D256$_{0.0}$}}
\def\Dnvh{\mbox{D1024$_{0.0}$}}

\def\mathnew{\mathsurround=0pt} 
\def\simov#1#2{\lower .5pt\vbox{\baselineskip0pt 
    \lineskip-.5pt\ialign{$\mathnew#1\hfil##\hfil$\crcr#2\crcr\sim\crcr}}}   
  
\def\simless{\mathrel{\mathpalette\simov <}}  
\def\'#1{\ifx#1i{\accent"13\i}\else{\accent"13#1}\fi}  
\def\eg{e.g.,}

\def\plotone#1{\centering \leavevmode
   \epsfxsize=\columnwidth \epsfbox{#1}}

\def\plotancho#1{\includegraphics[width=18cm]{#1}}

\begin{document}

\title{On  the Structure of Dark Matter Halos at the Damping Scale of 
the Power Spectrum with and without Relict Velocities } 
\author{Pedro Col\'in} 
\affil{Centro de Radiostronom\'ia y Astrof\'isica, Universidad Nacional Aut\'onoma de M\'exico, 
Apartado Postal 72-3 (Xangari), 58089 Morelia, Michoac\'an, Mexico}

\author{Octavio Valenzuela and Vladimir Avila-Reese} 
\affil{Instituto de Astronom\'ia, Universidad Nacional Aut\'onoma  
de M\'exico, C.P. 04510, M\'exico, D.F., M\'exico}  

\keywords{cosmology:dark matter  --- galaxies:halos --- 
methods:N--body simulations}

\begin{abstract} 
We report  a series of high--resolution cosmological N-body  simulations 
designed to explore the
formation and properties of dark matter halos with masses close to the
damping scale of the primordial power spectrum of density fluctuations. 
We further investigate the
effect that the addition of a random  component, \vth, into the particle
velocity field has on the structure of halos.
We adopted  as a fiducial model the $\Lambda$ warm dark matter ($\Lambda$WDM)
cosmology with a non--thermal sterile neutrino  mass  of 0.5 keV. The filtering
mass corresponds then to $M_f = 2.6 \times 10^{12} \msunh$. 
Halos of masses close to $M_f$ were simulated with several million of particles.  
The results show that, on one hand, the inner density slope of these
halos (at radii $\simless 0.02$ the virial radius \rvir) is
systematically steeper than the one corresponding to the
Navarro-Frenk-White (NFW) fit or to the cold dark matter
counterpart. On the other hand, the overall density profile (radii
larger than 0.02\ \rvir) is less
curved and less concentrated than
the NFW fit, with an outer slope shallower than -3.
For simulations with \vth, the inner halo
density profiles flatten significantly at radii smaller than 2--3
\kpch~ ($\simless 0.010\rvir-0.015\rvir$).
A constant density core is not detected in our simulations, with the 
exception of one halo for which the flat core radius is
$\approx 1 \kpch$. Nevertheless, if ``cored'' density profiles
are used to fit the halo profiles, the inferred core radii are 
$\approx (0.1-0.8)\kpch$, in rough agreement
with theoretical predictions based on phase--space constrains, and 
on dynamical models of warm gravitational collapse. 
A reduction of \vth~ by a factor of 3 produces
a modest decrease in core radii, less than a factor of 1.5.
We discuss the extension of our results into several contexts, for example, to
the structure of the cold DM micro--halos at the damping scale of this model.

\end{abstract}     

 
\section{Introduction} 

\label{intro}

The nature of dark matter (DM) is one of the most intriguing 
and fundamental problems in cosmology and particle physics.  
The standard hypothesis assumes that dark matter is made of  
non--baryonic collisionless elemental particles that become  
non--relativistic very early in the history of the Universe (cold). 
This minimal scenario, named Cold Dark Matter (CDM), has successfully 
explained the observed structure of the universe at large scales, like 
the two--point correlation function of galaxies and the Cosmic Microwave Background
Radiation (CMBR) anisotropies \citep[see for recent results][]
{Springel06,spergel2006}. Confrontation of model predictions with 
obervations turns out to be more complicated at galactic scales, because 
non--linear dynamics and baryonic processes may distort considerably the 
underlaying DM distribution. Thus, the predictions of the CDM scenario 
and its variants at the scale of galaxies, are an active subject of study.    

Two of the most controversial CDM predictions are: (i) the large abundance 
of subhalos in galaxy--sized halos \citep{kwg1993},  
and (ii) the cuspy inner density profile of dark halos \citep[\eg][]{ navarro2004, diemandprof}.   
Based on comparison with observations, it has been suggested  that both 
predictions may indicate a flaw of the CDM scenario 
\citep[][Gentile et al. 2005,2007]{klypin1999a, moore1999a,moore94,deblok01}.
However, these comparisons might be biased by astrophysical processes in action 
during galaxy assembly and evolution, 
as for example the inhibition of star formation in small subhalos 
\citep[][]{bullock00,benson02, governato07} or the halo core expansion 
due to energy or angular momentum transfer from dark/baryonic structures
\citep[][]{Ma2004,elzant04,weinbergkatz07}, but see also \citep{colin2006,ceverino07,sellwood07}.
It has also been  showed that the disagreements may be 
a consequence of systematics in the observational 
inferences, \citep[\eg][]{simon-geh07,rhee04, hayashi06,valenzuela07}. 

It is also possible that slight modifications to the CDM particle 
properties could solve or ameloriate the mentioned potential 
problems if they persist \citep[\eg][]{selfinteract1, selfinteract2}. 
As the precision of observations
and the control of systematics improve, the confrontation with 
model predictions opens a valuable window for constraining  the dark matter 
properties. Among the ``slight'' modifications with the CDM scenario
is the introduction of warm particles, instead of cold ones (Warm Dark Matter, 
hereafter WDM). WDM implies two more degrees of freedom in comparison to CDM: 
(i) a damping (filtering) of the power spectrum at some intermediate scale, 
\fsl, due to the relativistic free streaming, and (ii) some primordial random 
velocity in the dark particles,  \vth~ \citep[for CDM, the mass corresponding
to \fsl, $M_f$, is comparable to planet masses, e.g.,][and $\vth~ \approx 0$]
{diemandN05,profumo06}.

Previous numerical studies have shown that WDM can be
very effective in reducing the amount of (sub)structure below the
filtering mass $M_f$ \citep[\eg][]{cav2000,acv2001, bode2001, knebe02}. 
In addition, both the peculiar dynamical formation history
of these halos and its random \vth~ can also impose an upper 
limit on the phase space density, potentially producing 
an observable core of constant density \citep{hd2000, acv2001}.
If \vth\ has a thermal origin its amplitude is linked directly to the mass 
of the WDM particle \citep{hd2000}; notice, however, that the amplitude of 
the random velocities may depend on other physical factors not
directly related  to the particle mass. This is the case, for instance, 
of gravitinos produced non--thermally by late decays of the Next to Lighest 
Supersymmetric Particle (NLSP) \citep[][see for a recent 
review Steffen 2006]{frt2003, strigari2007}.  Thus, an exploration 
of the effect of a random velocity component independently of the dark matter 
particle mass seems to be necessary. 

Predictions of the core radius in WDM halos have been computed assuming
a King profile \citep{hd2000} or a subclass of the \citet[][]{zhao96} 
profiles \citep[][herafter S2006]{strigari2006}.  It is  still unknown which 
estimates give the more accurate value, yet these predictions are necessary 
for comparison with observations \citep[see e.g., S2006;][]{gilmore07}. 
Unfortunately, the predicted core radii for masses of the most popular 
WDM particle, the sterile neutrino, 
allowed by observational constraints ($\ga 2\ \kev$, see \S 5 for references)
are below the resolved scales in current simulations. 

The structure of WDM halos of scales below the filtering radius \fsl~ 
might be different from their CDM counterparts not only in the central
parts but also in their overall mass distribution. This is somehow
expected because the assembly history of these halos is different
from the hierarchical one. Besides, they form later and have  
concentrations lower than those ones derived for CDM halos
\citep[][]{acv2001,bode2001,knebe02}. However, it is controversial whether 
the shape of the density profile differs systematically from the corresponding 
CDM one \citep[see for different results e.g.,][]{jhs1999,moore1999b,knebe03}. 
On the other hand, WDM halos of mass close to $M_f$ (with \vth~ set to 0) can be thought of
as scaled-up versions of the 
first microhalos in a CDM cosmology. The earliest collapse of CDM
microhalos is a subject of considerable current interest 
 \citep[e.g.,][]{diemandN05,gao05}.

In this paper we explore by means of numerical simulations the 
two questions mentioned above: the overall structure of dark halos 
with masses close or just below the cutoff mass in the 
power spectrum of fluctuations, and the inner density profile of these 
halos when a random velocity is added to the particles.
For our numerical study we  use the truncated power spectrum corresponding
to a non--thermal sterile neutrino of $\mnu=0.5$ keV
($M_f = 2.6 \times 10^{12}\ \msunh$). We
initially neglect  the particle random velocity (\vth=0), and later we will consider two 
values of \vth~ that cover the range of velocities corresponding to thermal 
and non--thermal $\mnu=0.5$ keV WDM particles. It is important to 
remark that our goal is not   to study a specific  WDM model but to explore in general
the effects on halos of the power spectrum truncation and the addition of
random velocities to the particles.

The structure of the paper is as follows. In \S 2 we describe
the cosmological model that we use for our investigation :
a WDM model with a filtering radius at the scale of Milky Way size halos. 
Halos of these scales were resimulated with higher resolutions, first
without adding the corresponding random velocity, and then with 
adding this velocity component to the particles. Details 
of the numerical simulations carried 
out in this paper are given in \S 3. The results from our different 
simulations are presented in \S 4. Section \S 5 is devoted to a 
discussion of the results and their implications. Finally, in  
\S 6 we present the main conclusions of the paper.

\section{The cosmological model}

The general cosmological background that we use
for our numerical simulations corresponds to the 
popular flat low--density model with $\ome = 0.3$, 
$\omel = 0.7$ and $h = 0.7$ (the Hubble constant 
in units of $100~\hubble$).

For the experiments designed to explore the 
structure of dark matter halos with masses close
to $M_f$, we adopt an initial power spectrum 
corresponding to a non--thermal sterile neutrino of 
$0.5$ keV. Even if this WDM model is ruled out by observations
(see \S 5 for references), it is however adequate for
the purposes stated in the Introduction.  As is shown 
below, the filtering mass $M_f$ corresponding to 
this WDM particle is of the order of Milky Way--size halos, 
namely the halos that we are able to follow with high resolution
in a cosmological simulation. The high resolution of the   simulations avoids 
that the formation of halos with a  mass  scale  near to  $M_f$ 
will be  dominated by discreteness 
effects (see \S \ref{discr}). Besides, for the resolution
that we attain, we expect to resolve the inner regions
where a flattening in the halo is predicted density profile is predicted
for  the case when  \vth is introduced.

We use here the transfer function $T_s$ for the non--thermal sterile
neutrino derived in \citet{kev2006a}. The WDM power spectrum is then given by
\begin{equation}
P_{WDM} (k) = T^2_{s} (k) P_{CDM} (k),
\label{pswdm}
\end{equation}
\noindent where 
\begin{equation}
T_s (k) = \left[ 1 + (\alpha k )^\nu \right]^{-\mu},
\end{equation} 
\noindent and $P_{CDM}$ is the CDM power spectrum given by 
\citet{kh1997}. This fit is
in excellent agreement, at the scales of interest, with the power spectrum 
obtained with {\it linger}, which is contained inside the 
{\it cosmics} package\footnote{http://web.mit.edu/edbert/cosmics-1.04.tar.gz}.
The parameter $\alpha$ is related to the mass of  
the sterile neutrino, $\Omega_{DM}$ and $h$ through 
\begin{equation}  
\alpha = a \left( \frac{m_s}{1 \kev} \right)^b \left( \frac{\Omega_{DM}}{0.26} \right)^c
\left( \frac{h}{0.7} \right)^d \mpch,
\end{equation} 
where $a= 0.189$, $b=-0.858$, $c=-0.136$, $d=0.692$, $\nu = 2.25$, and $\mu =
3.08$. The power spectrum is normalized to $\sige = 0.8$, a value close to that
estimated from the third-release of WMAP
\citep{spergel2006}.
Here \sige\ is the rms of mass fluctations estimated with the top-hat
window of radius $8 \mpch$.

As in \citet{acv2001}, we have defined the free--streaming (damping) 
wavenumber, $k_f$, as the $k$ for which the WDM transfer 
function $T^2_s(k)$ decreased to 0.5, and compute the corresponding 
filtering mass in the linear power spectrum as
\begin{equation}
M_f = \frac{4 \pi}{3} \bar\rho \left(\frac{\lambda_f}{2} \right)^3,
\end{equation}
where $\bar\rho$ is the present--day mean density of the universe.
The filtering wavelength $\fsl = 2\pi/k_f$
of a non--thermal sterile neutrino of $0.5\ \kev$ is $3.9\ \mpch$, which 
corresponds to a filtering mass $M_f = 2.6 \times 10^{12}\ \msunh$.
The random component was linearly added
to the velocities  calculated with  the Zel'dovich approximation at the onset of the simulation.

A number of Milky Way--size halos are simulated with the WDM power
spectrum given by eq. (\ref{pswdm}) and with \vth=0. In this way we
isolate the effects of the power spectrum filtering on the structure
of the halos. Later, the same halos are resimulated but adding random 
velocities to the particles.   
We approximate the shape of the particle phase space distribution    
function (DF)  with the corresponding thermal one; i.e, we use
a thermal equilibrium Fermi-Dirac DF. This is a good approximation for
a non--thermal sterile neutrino of 0.5 keV \citep{kev2006a}. Further, as a 
first approximation, we assume that the amplitude of the DF of our non--thermal
sterile neutrinos is equal to the one corresponding to their thermal 
partners with \mnu=0.5 keV \citep{hd2000}. With this  assumption for DF shape and amplitude,  
the neutrino random  velocity at $z = 40$, the initial redshift in our simulations, 
is $\approx 4\ \kms$. 

The particle velocity dispersion in the linear 
regime decreases adiabatically with the expansion. The convention is to define 
\vth~ in terms of the value extrapolated to the present epoch,    
$\vth(z)=\vth(0)(1+z)$; therefore, for our case  $\vth(0)\approx 0.1$ \kms.
We should note that the rms velocity amplitudes of sterile neutrinos
and thermal particles of the same mass can not be the same, but larger
because the former become non--relativistic later than the latter. 
A rough calculation shows that for the 0.5 keV sterile neutrino, \vth~
should be approximately three times higher than for a 0.5 keV thermal
WDM particle, i.e.  $\vth(0)\approx 0.3$ \kms. We resimulate halos
with both values of the random velocity, $\vth(0)\approx 0.1$ and 0.3 \kms,
having in mind that our main goal is not to study a particular WDM
model but to explore the structure of the dark halos with different
initial conditions  .

\section{Numerical Simulations}

A series of high resolution simulations of Milky Way size galactic halos 
are performed using the Adaptive Refinement Tree (ART)
N-body code \citep{kkk1997} in its multiple mass scheme \citep{kkbp2001}. 
The ART code achieves high spatial resolution by refining the root grid in
high-density regions with an automated refinement algorithm.

In all of our experiments we first run a low--mass resolution (LMR) simulation
of a box of $L_{box}=10$\mpch~ on a side and $64^3$ or 
$128^3$ particles. We then select a spherical region centered on a Milky 
Way size halo of radius $\sim 3$ times the virial 
radius\footnote{This radius is defined as 
the radius at which the mean overdensity is equal to the predicted by a top-hat 
spherical collapse which is 337 for our selected cosmological model.} of 
the chosen halo. The Lagrangian region corresponding to the $z = 0$ spherical
volume is identified at $z= 40$ and resampled with additional small--scale waves
\citep{kkbp2001}. Halo names (first column in Table 1) are denoted by a capital 
letter followed by the effective number of particles that were used 
in the high resolution zone and a subindex that indicates the value of 
$\vth(0)$ (in \kms; reported in column 2)\footnote{The random velocity component
was  linearly added to the velocities correspondingt to  the Zeldovich approximation 
at the onset of the simulation.}.
The mass per particle, $m_p$, in 
the high--resolution region is given in column (5) while in column (6) we give 
the virial mass of the halo, \mvir. The  latter along with $m_p$ 
can be used to compute the number 
of particles inside a halo. The global expansion timestep $\Delta a_0$, 
and the formal spatial resolution $h_{for}$ --measured by the size of a  
cell in the finest refinement grid-- are given in columns (3) and (4).  
ART integrates the equations using the expansion factor $a$ as the time  
variable such that at $z=0$, $a=1$.  

\begin{deluxetable*}{cccccccccccccc}
\tablecaption{Model and Halo Parameters$^{a}$}
\tablewidth{0pt}
\tablehead{\colhead{name tag} & \colhead{$\vth$} & 
\colhead{$\Delta a_0$} & \colhead{$h_{for}$} &\colhead{$m_p$} & \colhead{M$_{v}$} &
\colhead{\vmax} & \colhead{R$_v$} & \colhead{\cq} & \colhead{\cv} & \colhead{Lg$\bar{\rho}_{1\%}$} &
\colhead{r$_c$} & \colhead{r$_c^{S+}$} & \colhead{r$_{c,30}^{S+}$} \\
 & km/s & $10^{-3}$ & kpc/h & $10^6\msun/h$ & $10^{12}\msun/h$ & km/s & kpc/h &  &  &   
 [$h^2\msun/pc^3$] & kpc/h & kpc/h & kpc/h   \\
(1) & (2) & (3) & (4) & (5) & (6) & (7) & (8) & (9) & (10) & (11) & (12) & (13) & (14)}
\startdata
\Anv     & 0.0 & 1.0 & 0.305  &  4.96  &  2.84 & 224.1 & 286 & 5.6 & 8.3 & -1.44 & -- & -- & --\\
\Av      & 0.1 & 1.0 & 0.305  &  4.96  &  2.37 & 218.2 & 228 & 5.0 & 5.9 & -1.44 & 0.082 & 1.151 & 0.572\\
\Bnv     & 0.0 & 0.5 & 0.305  &  4.96  &  4.36 & 253.2 & 330 & 5.1 & 7.7 & -1.46 & -- & -- & --\\
\Bv      & 0.1 & 0.5 & 0.305  &  4.96  &  4.35 & 250.8 & 330 & 5.0 & 6.7 & -1.54 & 0.098 & 1.380 & 0.565\\
\Cnv     & 0.0 & 0.5 & 0.152  &  0.62  &  1.29 & 160.7 & 222 & 4.2 & 6.4 & -1.53 & -- & -- & --\\
\Cv      & 0.1 & 0.5 & 0.152  &  0.62  &  1.26 & 158.9 & 220 & 4.3 & 4.7$^b$ & -1.79 & 0.093 & 1.265 & 0.689\\
\Cvv     & 0.3 & 0.5 & 0.152  &  0.62  &  1.28 & 158.5 & 221 & 4.3 & 3.6$^b$ & -1.99 & 0.121 & 1.527 & 0.827\\
\Dnv     & 0.0 & 0.5 & 0.152  &  0.62  &  1.28 & 160.1 & 221 & 4.3 & 6.1 & -1.54 & -- & -- & --\\
\Dv      & 0.1 & 0.5 & 0.152  &  0.62  &  1.24 & 160.9 & 219 & 4.6 & 4.2$^b$ & -1.89 & 0.106 & 1.389 & 0.816\\
\Dvb     & 0.1 & 0.5 & 0.152  &  0.62  &  1.25 & 159.6 & 219 & 4.3 & 4.8$^b$ & -1.77 & 0.091 & 1.246 & 0.598\\
\Dvvb    & 0.3 & 0.5 & 0.152  &  0.62  &  1.26 & 160.4 & 220 & 4.3 & 3.7 & -1.96 & 0.118 & 1.498 & 0.792\\
\Env     & 0.0 & 1.0 & 0.152  &  0.62  &  2.13 & 197.3 & 262 & 4.9 & 6.4 & -1.57 & -- & --& -- \\
\Ev      & 0.1 & 1.0 & 0.152  &  0.62  &  2.06 & 192.3 & 259 & 4.7 & 5.7$^b$ & -1.66 & 0.092 & 1.287 & 0.657\\
\Evv     & 0.3 & 1.0 & 0.152  &  0.62  &  2.14 & 197.0 & 262 & 4.9 & 5.3$^b$ & -1.71 & 0.101 & 1.380 & 0.519\\
\enddata 
\tablenotetext{a}{All the halos presented in this Table were resimulated from a $L_{\rm box}$ 10\mpch~ box simulation.}
\tablenotetext{b}{Note that the NFW function does not provide a good fit to the density profiles of halos with $\vth>0$. However, for completeness, we report here the value of \cv~ obtained from the fit.}
\end{deluxetable*}

\begin{figure}[htb!] 
\plotone{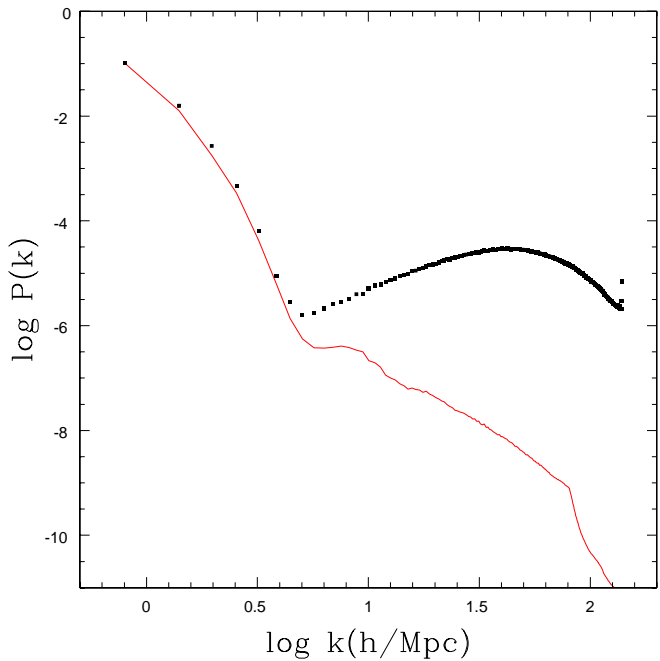}
\caption[fig:pk]{Comparison of the power spectra measured at $z=40$ for the
simulation started at $z=40$ (solid line) and the one started at $z=100$ (squares).
The longest plotted wavelength is $L_{box}$ while the highest frequency is 
$ \left( 2\pi/L_{box} \right)256$.
\label{fig:pk} }
\end{figure}

We started the simulations at $z = 40$ because the power at the 
Nyquist frequency at this redshift is in the linear regime. 
Note that spurious noise can influence the evolution of simulations that
include \vth\ if they start too early. The noise might be particularly important 
when the generated \vth\ have amplitudes comparable to those of the Zel'dovich
peculiar velocities. In order to show this, we started a simulation at $z = 100$ 
and evolved it up to $z = 40$. At $z = 100$ the \vth\ velocities
are on average 2.5 times greater than at $z = 40$.
Figure \ref{fig:pk} shows the measured initial power spectrum of the
simulation started at $z=40$ (solid line) and the power spectrum
for the simulation started at $z=100$ and measured at $z = 40$ (squares).  
The latter simulation developed spurious power at frequencies higher than 
about 0.6 $h/$Mpc. We  run another  simulation started at $z=100$, but with 
no addition of a relic velocity component. In this case
we did not detect the evolution of spurious power by $z =40$.
In order to make sure that this spurious noise does not appear when
the simulation is started at $z=40$ we repeated the experiment but now 
the onset of the simulation is set at $z=40$ and the power spectra are
measured at $z=20$.
Unlike the previous case, no differences between the power spectra were detected.
In other words, as far as the initial power spectrum is concerned it does not
matter if the simulation is started at $z=40$ o $z=20$.

Concern may arise about the structure of halos simulated in
a relatively small $L_{box}=10$\mpch\ box, specially in a WDM cosmology, where
there is a scale below which the power spectrum exponentially
drops to zero. \citet{acv2001} discussed this potential issue and
concluded that in order to be confident about the simulated
halo structure, a box size greater than the filtering length \fsl~ should be used. 
For our simulations $L_{box}$ is 2.5 larger than \fsl. In any case, we also 
experimented with other box sizes, namely, 15\mpch\ and 20\mpch\
(not shown in the Tables), for the $\vth = 0$ case,
and found results similar to those reported here for the 10\mpch\ box.

\begin{deluxetable*}{ccccccc}
\tablecolumns{7}
\tablewidth{0cm}
\tablecaption{High Resolution Halo D Parameters}
\tablehead{\colhead{$L_{box}$} & \colhead{name tag} & \colhead{$\vth$} & 
\colhead{timestep} & \colhead{resolution} &\colhead{$m_p$} & \colhead{M$_{vir}$} \\
($\mpch$) &  &  & ($10^{-3}$) & ($\kpch$) & ($\msunh$) & ($10^{12}\ \msunh$) \\
(1) & (2) & (3) & (4) & (5)& (6) & (7)}
\startdata
 10 & \Dnvh      & off & 0.5 & 0.040  &  $7.75 \times 10^4$  &  1.28 \\
 10 & \Dnv       & off & 0.5 & 0.152  &  $6.20 \times 10^5$  &  1.27 \\
 10 & \Dnvl      & off & 0.5 & 0.305  &  $4.96 \times 10^6$  &  1.27 \\
 10 & \Dnvll     & off & 0.5 & 0.610  &  $3.97 \times 10^7$  &  1.25 \\
\enddata
\end{deluxetable*}

The bound density maxima (BDM) group finding algorithm \citep{klypin1999b},
or a variant of it \citep{kravtsov2004}, is used to locate the halos in the 
simulations and to generate their density profiles. The BDM finds positions 
of local maxima in the density field smoothed at the scale of interest and 
applies physically motivated criteria to test whether a group of particles 
is a gravitationally bound halo. 

Aside from those halos shown in Table 1, for the halo D with \vth=0 we have also 
run a very high resolution simulation with about 31 million particles in the high 
resolution zone.  This halo was taken from the same 10 \mpch\ box run but 
with $128^3$ particles in the LMR mode.
As far as we know this is the highest resolution simulation of a halo 
run in a WDM cosmology. The same halo was also simulated with less resolution
for a convergence test. The parameters of the sequence of halos D are resumed in 
Table 2.  

The simulations presented here differ in several aspects from previous WDM
simulations. First, it should be emphasized that our aim rather than discussing
a specific WDM model is to explore the influence of the truncation of the
power spectrum and/or the addition of random velocities upon the structure of 
dark halos of masses close to the truncation scale.  For this aim we need to simulate 
(a) halos with very high resolution, and (b) halos with masses close to $M_f$. The halos 
simulated in \citet{acv2001} had several times less particles than the best--resolved 
halos presented here and the aims in that paper focused in exploring general halo 
properties for a concrete WDM model. Other papers aimed to study the 
properties of WDM halos 
\citep{bode2001,knebe02,busha07} focused more in the statistical aspects than in 
details of the inner halo structure; therefore, the halos in these papers had resolutions 
much lower than those attained here. The properties of the WDM halos simulated here are
in general agreement with previous findings; for example, their concentrations are
systematically lower \citep{acv2001,eke01,bode2001} and they form later 
\citep{knebe02,busha07} than the corresponding $\Lambda$CDM halos.   

\subsection{Discreteness effects} 
\label{discr}

One of the motivations of this paper is to investigate the structure of well
resolved halos with masses close or below the damping (truncation)
scale in the power spectrum, $M_f$. 
The origin of these halos is controversial.  
Halos with masses close to $M_f$ (truncation halos) 
could be formed by a quasi--monolithic
collapse of filaments of size $\sim \fsl$ \citep[e.g.,][]{acv2001}.
They could also be just the result of an incomplete collapse, highly deviated
from the spherical--symmetric case, of originally larger  
structures assembled hierarchically \citep{busha07}. On the other hand, it has been
suggested that halos
with masses considerably less than $M_f$  form by fragmentation
of the shrinking filaments of size $\sim \fsl$ \citep[e.g.,][]{Valinia97,
acv2001,bode2001,gotz03,knebe03}. However, it is also known that the filaments 
in Hot Dark Matter simulations that start from a cubic lattice break up into 
regularly spaced clumps, which reflect the initial grid pattern. Therefore, some of 
these halos seen in WDM 
simulations could be spurious, product of discreteness effects. Recently, 
\citet{ww2007} have shown that this artifact is present even for a glass--like 
initial particle load \citep[][]{white96}.   

As \citet[][]{ww2007} show, halos of masses smaller than a given effective fraction 
of $M_f$, which depends on the resolution of the simulation, will be spurious.    
We selected the WDM model (\S 2) and the number of particles in the simulations 
in such a way that the halos studied here can neiher be fake nor affected by 
discreteness  effects.  Following \citet{ww2007}, in our model with $1024^{3}$ 
effective number 
of particles, only structures with  masses lower than 1.3 $\times 10^{9}\msunh$ are 
candidate to be spurious. 
For the models with $512^{3}$ and $128^{3}$ effective number of particles, the 
masses of structures  
triggered by the initial grid spacing are $<2.6 \times 10^{9}\msunh$ and 
1.04 $\times 10^{10}\msunh$, respectively.  For our WDM model, the first 
halos to form have masses  $\ga 10^{12} \msunh$, which are far from the mentioned 
effective resolution limits.  
\begin{figure*}[htb!] 
\vspace{7.5cm}
\hspace{15.3cm}
\includegraphics{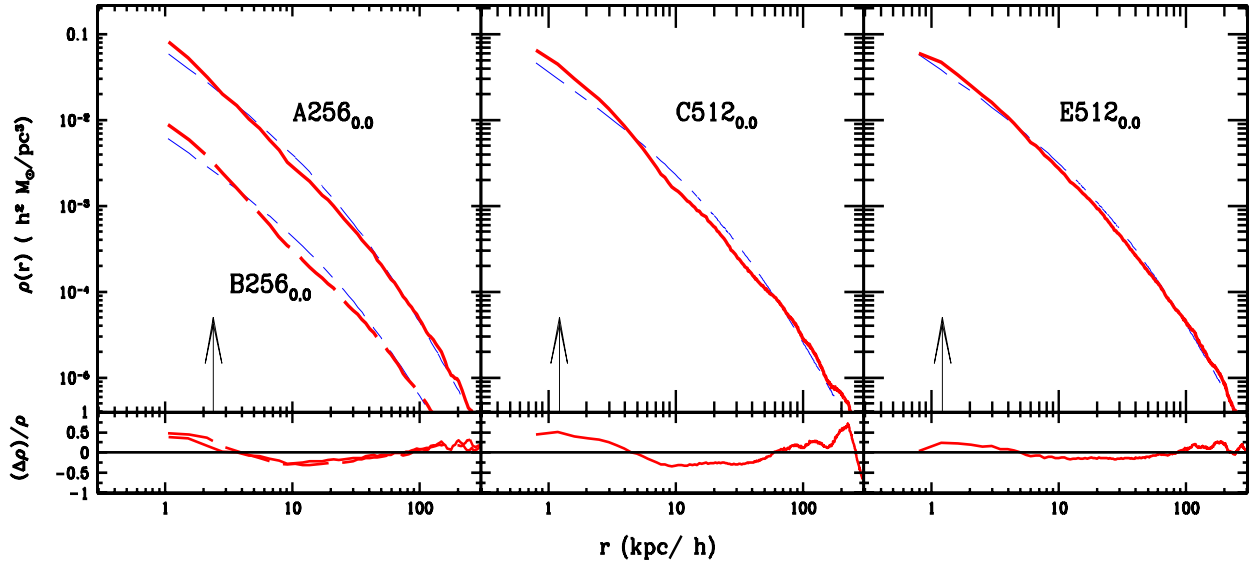}
\caption[fig:perfnov]{{\it Upper panels:} Density profiles of halos formed at the
scale of damping (truncation). From the left, the first two panels present 
the profiles (thick lines) of two halos each one; the lower profiles
were shifted by $-1$ in the log for presentation purpouses.  Thin short--long 
dashed lines show the corresponding NFW fits. The third panel shows halo 
K simulated with four different resolutions. The NFW fits are showed only
for the two highest resolutions.  The innermost radius of each profile
corresponds to $\approx 1.3 \ h_4$ and arrows in each panel indicate the radius $h_8$
(see text).   {\it Lower  panels:}  Fractional residuals of the 
NFW fit. We observe that all our halos are denser and steeper than the NFW 
model at least until $h_8$.
\label{fig:perfnov} }  
\end{figure*}  

\section{Results}

We first present the results of our WDM simulations with $\vth=0$.  
In this way, we explore the inner structure 
of halos close to the truncation mass $M_f$, which could be scaled-up 
versions of  the first CDM microhalos (of masses $\sim 10^{-5}\msun$ for a neutralino  
mass of 100 GeV).  Afterwards we present the results of the same simulations  
but introducing random velocities to the particles with two different amplitudes,
$\vth(0)=0.1$ and 0.3 \kms~ (see \S 2). The main goal of the last simulations  
is to explore the predicted flattening in the halo inner density profile
produced by the addition of a random component to the particle 
velocities(see section \ref{intro}).     
Table 1 resumes the main properties of the resimulated halos, 
which were selected to have masses close to the truncation mass
of the initial power spectrum.

\subsection{The Structure of Halos at the scale of Damping} 

\begin{figure}[htb!] 
\plotone{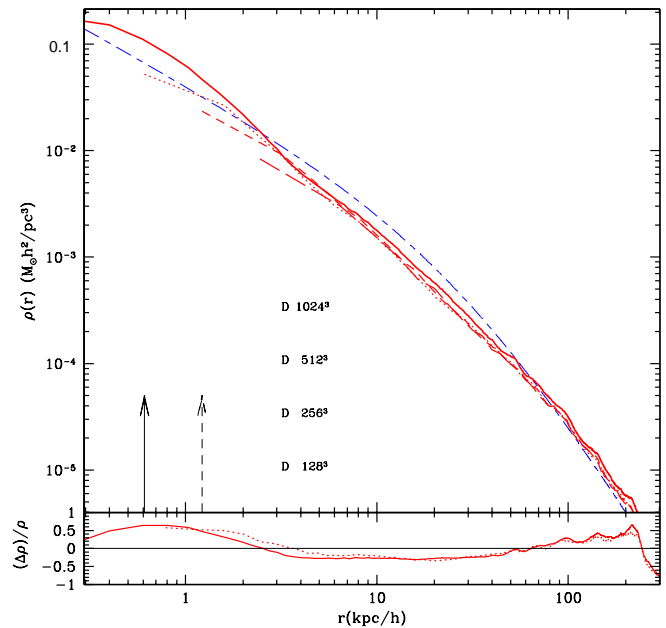}
\caption[fig:haloD]{{\it Upper panel:} Density profile for halo D. This halo was simulated 
with four different resolutions, separated each by a factor of eight in particle mass
(see Table 2). Shown with a short-long-dashed curve is the NFW fit to the density
profile for the most well resolved case (solid line) while arrows indicate the position of the radii
$h_8$ for the two highest resolution simulations.
\label{haloD} }  
\end{figure}

Figures \ref{fig:perfnov} and \ref{haloD} show the spherically averaged density profiles 
measured for the halos of masses around the power--spectrum filtering 
mass, $M_f$, and with \vth=0.  
The first panel of Fig. \ref{fig:perfnov} shows halos \Anv~ and \Bnv~ (the latter was 
shifted by $-1$ in the log); the second and third panels show halos
\Cnv\ and  \Env, respectively.
In Fig. \ref{haloD}, the same halo $D$ but simulated with 4 
different resolutions is shown.  In both Figs., the thin dashed lines are the 
best Navarro--White--Frenk \citep[NFW,][]{nfw} fits to the showed density
profiles. In the lower panels, the residuals of the 
measured density profile and the NFW fit are plotted with the same line 
coding as in the corresponding upper panels.

Halos \Anv~ and \Bnv~ in the left panel and \Cnv~ and \Env~ in the second
and third panels of Fig. \ref{fig:perfnov}, were simulated with formal spatial resolutions 
$h_{\rm for}= 0.305\kpch$ and 0.152\kpch, respectively (see Table 1). 
Previous convergence studies for CDM halos simulated with the ART code have 
shown that the innermost halo density is reliable only for radii larger than 
four times $h_{\rm for}$ and containing  more than 200 particles 
\citep{kkbp2001}.  For all the density  profiles shown in Figs.  \ref{fig:perfnov}
and \ref{haloD}, the innermost plotted point corresponds to radii larger than 
$h_4= 4\times h_{\rm for}$ by $\sim 30\%$ and they contain more than 200 
particles.  The convergence analysis that we have carried out for our WDM 
halos suggests that instead of $h_4$, the innermost radius should
be close to $8\times h_{\rm for}$, $h_8$.  

Figure \ref{haloD} compares the density profiles of halo D, which was re--simulated 
with four different resolutions separated each by a factor of eight in the particle 
mass (see Table 2).  As can be seen, convergence is achieved at about $h_8$.  
The arrows in Figs. \ref{fig:perfnov} and \ref{haloD} indicate $h_8$ for the 
corresponding simulations. 
In Fig.\ref{haloD} , the solid--line arrow is for the highest--resolution 
simulation ($1024^3 $ effective number of particles), while the dashed--line arrow is for the  $512^3$ simulation; 
the NFW fit is shown only for these two cases.   

The inner density profiles  of all halos simulated here are 
systematically steeper than the corresponding fitted NFW law.  The slopes of the 
profiles at $r\approx 1\%$ the virial radii, \rvir, span a range from 
$-1.4$ to $-1.6$.  For comparison, the slope of the density profile of a typical 
LCDM halo of $2\times 10^{12}\msunh$ at $0.01$\rvir~ is $\approx -1.2$ 
\citep[a NFW profile was used with the corresponding concentration given by][and
re-scaled to $\sigma_8=0.8$]{bullock01}. 
At radii smaller than $0.01$\rvir, the slopes tend to become shallower but the
halos are still denser and slopes steeper than the corresponding  
NFW fit up to the resolution limits 
($\approx h_8$, see Figs. \ref{fig:perfnov} and \ref{haloD}). 

The overall density profile shapes of our halos are also  somehow different
from the NFW function.  For radii larger than $0.02$\rvir, the  profiles tend to be in 
general slightly less curved than the NFW model. This is why the residuals shown in Figs.  
\ref{fig:perfnov} and \ref{haloD} indicate a systematical defect at intermedium radii 
and then an excess at the outer radii. The outer slopes are $> -3$. 

In summary, the density profiles of the halos simulated here,
with masses close or below the filtering mass, have a shape
slightly flatter than the NFW law for radii  $>0.02$\rvir~ and slopes significantly 
steeper at radii $<0.02$\rvir.  Nevertheless, each  profile is different. The profile 
of  halo \Env~ has minimal deviations from the NFW function, while the profile of 
halo \Dnv~ significantly deviates from this function.  

Table 1 details the main properties of the halos studied here.  In columns 6 to 14 are 
reported respectively the virial mass, \mvir, the maximum circular velocity, \vmax, the 
\cq~ and NFW concentration parameters, the average density within 1\% the virial
radius, and three core radii estimated by different criteria (the latter quantities 
apply only to halos simulated with \vth, see \S\S 4.2). The NFW and \cq~ 
concentrations are defined respectively as the ratios between \rvir~ and the NFW scale 
radius, and between \rvir~ and the radius where 1/5th of \mvir~ is contained 
\citep{AFKK99}.  As found in previous results, halos of masses below the truncation 
mass in the power spectrum
tend to be less concentrated than \lcdm~ halos of similar masses \citep{acv2001,bode2001,
knebe02}. We have simulated some $\Lambda$CDM ($\sigma_8=0.8$) halos of masses 
$\approx 2\times 10^{12}\ \msunh$ and measured NFW and \cq~ concentrations around 
8--15 and 6.0--11.0, respectively, to be compared with the values given in Table 1 for
the \vth=0 cases.  

\subsection{The inner structure of halos simulated with rms
velocity added to the particles} 
\label{veff}

\begin{figure*} 
\plotancho{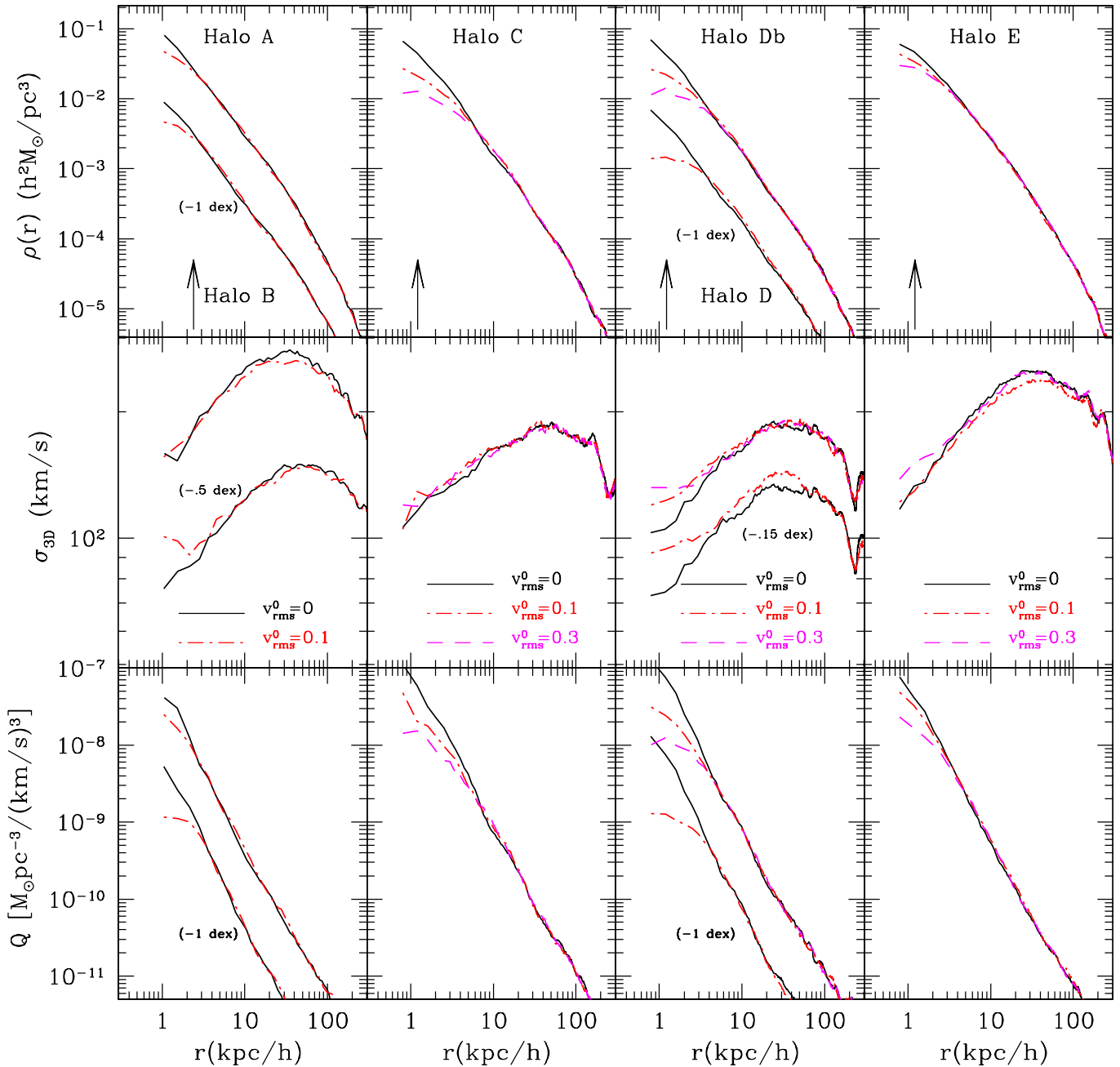}
\caption[fig:panel]{{\it Upper panels:} Comparing the density profiles of simulated
halos without (solid lines) and with (dot--dashed lines) adding \vth~ to the 
particles. In the left panel, halos B and B$_v$ have been shifted by $-1$ in the log. 
The innermost plotted radii are at $\approx 1.3\ h_4$ and the arrows indicate the corresponding 
$h_8$ radii.  For halo D, the profile of the $1024^3$ particles simulation (\vth=0)
is also plotted (dotted line).  {\it Medium panels:} The corresponding 3--D velocity 
dispersion profiles of the halos shown in the upper panels. Halos B  and B$_v$  have 
been shifted by $-0.5$ in the log.   {\it Lower panels:} Coarse--grained phase--space
density profiles corresponding to the halos shown in upper and lower panels.
\label{panel}} 
\end{figure*}   

We have rerun the halos presented in the previous 
sub--section but now introducing a random velocity component, \vth.
As explained in \S 2, the particle DF used corresponds to a Fermi--Dirac
function. Regarding the \vth~ amplitude, we use two values: $\vth(0)=0.1$ 
and 0.3 \kms. These values are for WDM particles of thermal origin
and for non--thermal sterile neutrinos, respectively, in both
cases with \mnu=0.5 keV.  Our goal is to explore whether the inner 
structure of the halos becomes affected significantly or not by adding 
\vth. 
 
We first present results for the case $\vth(0)=0.1$ \kms, and then
explore how the inner halo structures change when $\vth(0)$ is increased 
from 0.1 to 0.3 \kms, a more appropiate value for the 0.5 keV 
sterile neutrino model used to generate the initial power spectrum of the
simulations (see \S 2). For halo D, we have simulated with $\vth > 0$
the $512^3$ case. Unfortunately, due to limitations in our computational 
resources, it was not possible to run the simulation with $1024^3$ particles 
for $\vth > 0$.

In Fig. \ref{panel} we present spherically--averaged profiles of different
properties for all the simulated halos without and with \vth~ added. In the first 
and third columns two halos (A and B, and Db and D, respectively) are presented
but the latter ones are down shifted for clarity.
The upper panels of Fig. \ref{panel} show the density profiles of our halos
without (black solid lines), and with $\vth(0)=0.1$ \kms~ (red dot--dashed line)
and 0.3 \kms~ (magenta dashed line).  
In the medium panels are plotted the corresponding three--dimensional velocity 
dispersion profiles, $\sigma_{\rm 3D}(r)$, and the lower panels show the 
corresponding coarse--grained phase--space density profiles, 
$Q(r)=\rho(r)/\sigma_{\rm 3D}^3$.  As in Fig. \ref{fig:perfnov}, the arrows 
indicate the strong resolution limit radius $h_8$ of the
simulations. For halo D (third column), the simulations with $\vth>0$ 
were carried out with $512^3$ particles and using two different random
seeds for the particle DF calculation (see below).

To some degree, all the simulated halos have been affected in their
inner regions by the injection of initial random velocities to their particles. 
For the less resolved halos A, B (left panels),  
the deviations at $\sim h_8$ of the inner density profiles from the profiles 
obtained  in the simulations with \vth=0 are yet marginal, but in the expected 
direction. For the halos C, D, Db, and E simulated with  $512^3$ particles, the deviations
down to $h_8$ are significant: the density profiles systematically flatten
with respect to the corresponding \vth=0 cases. The radii at which the density
profiles of halos with \vth~ start to deviate (flatten) from the ones without
\vth, are $\sim 0.010-0.015\rvir$, well above the resolution limit of 
the simulations.

In a second series of experiments, we have resimulated halos C, Db and E ($512^3$ 
particles) with \vth~ three times larger, i.e. $\vth(0)=0.3$ \kms. The profiles
corresponding to these simulations are plotted in Fig. \ref{panel} with
dashed (magenta) lines. The flattening of the inner density profiles is clearly
more pronounced than in the simulations with $\vth(0)=0.1$ \kms.

The innermost density profiles actually vary from halo to halo. Again, halo E is 
the less affected not only by the damping in the power spectrum,  but also by 
the injection of \vth, and halo D is the most affected by both effects (lower 
dot--dashed curve in the corresponding panel).  The latter 
actually shows a ``true'' flat core already at $h_8$. Since the core
of this halo is too different with respect to the other ones, we decided to
explore if large difference can be explained by a rare fluctuation in the random
procedure of particle velocity assignment. Thus, the same halo \Dv~ was resimulated 
with a different seed in the random number generator used to draw the particle 
velocities. The upper curves in the third column panels of Fig. \ref{panel} correspond 
to the profiles for this halo, called \Dv b. The inner density profile of this halo 
is  not too different from the profiles of the other halos, though it remains
as the flatest one among all the simulated halos with $\vth(0)=0.1$ \kms.

\begin{figure} 
\plotone{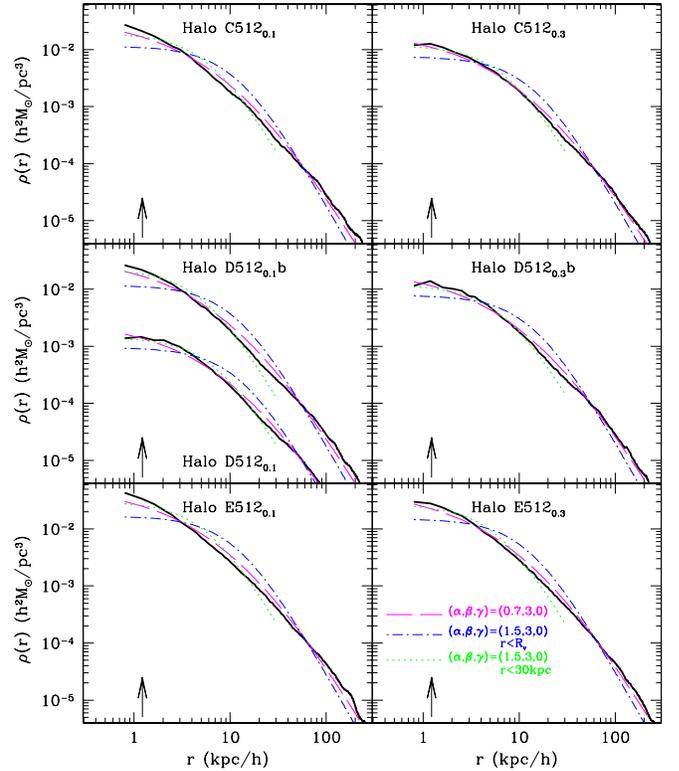}
\caption[fig:profwithv]{Density profiles of simulated ($512^3$ particles) halos 
with added \vth~ and fitted to different model profiles. Long--dashed line: 
$(\alpha,\beta,\gamma)=(0.7,3,0)$; dot--dashed line: the profile
proposed in S2006, $(\alpha,\beta,\gamma)=(1.5,3,0)$; dotted line:
the same S2006 profile but fitted only to the inner 30 kpc.
\label{profwithv}}
\end{figure}

In Fig. \ref{profwithv} we attempt to fit different functions to the density
profiles of halos C, D, and E ($512^3$ particles) with $\vth(0)=0.1$ and
0.3 \kms.  A general function
to describe density profiles of cosmic objects was proposed by \citet{zhao96}:
\begin{equation}
\rho(r)=\frac{\rho_0}{(r/r_0)^\gamma[1+(r/r_0)^\alpha]^{(\beta-\gamma)/\alpha}}
\label{zhao}
 \end{equation}
The NFW profile corresponds to $(\alpha,\beta,\gamma)=(1,3,1)$. This function does not
provide a good description for the profiles of our halos with $\vth>0$, in particular 
in the inner regions. We have fitted the halo profiles to the NFW in order to obtain
an estimate of the c$_{\rm NFW}$ concentration given in Table 1.

\citet{strigari2006} suggested a ``cored'' density profile in order to derive constraints 
on the size of a possible shallow core in the halo of the Fornax dwarf spheroidal galaxy 
\citep{goerdt06,sanchez06}.  This profile is described by eq. (\ref{zhao}) with
 $(\alpha,\beta,\gamma)=(1.5,3,0)$, and they define the core radius, \rc~  as the radius
where the inner slope, $g$, reaches the value of $-0.1$. Thus, 
$\rc = r_0/(-3/g -1)^{1/\alpha} \approx 0.1r_0$.  
The dot--dashed curves in Fig. \ref{profwithv} show the best fits using the
S2006 profile. As can be seen, this profile does not describe well the density profiles
of the WDM halos in the simulations. Note that $\alpha$ characterizes the sharpness of the 
change in logarithmic slope. As already seen in Figs. \ref{fig:perfnov} and \ref{panel}, the
profiles of our halos tend to be less curved than the usual NFW profile. Therefore,
values of $\alpha$ smaller than 1 should be used instead of larger than 1. We have obtained
a reasonable description of our WDM profiles with $(\alpha,\beta,\gamma)=(0.7,3,0)$.
The (magenta) dashed curves in Fig. \ref{profwithv} are the best fits with these
profiles. The core radius, as defined above, is in this case 
$\rc\approx 0.0064 r_0$. Finally, we have also 
tried fits to the S2006 function but taking into account only the central halo regions, 
up to $\approx 30$\kpch.
The fits are shown in  Fig. \ref{profwithv} with (green) dotted curves. The fit is specially 
good for the halo \Dv~ and those with \vth(0)=0.3 \kms.  Columns 12 to 14 in Table 
1 report the values of \rc~ obtained with the three different fits. 

As to the 3--D velocity dispersion profiles of the simulated halos, they do not differ 
significantly between the cases with and without \vth, the exception being halo D. In
the innermost regions, $\sigma_{\rm 3D}(r)$ is similar or higher for halos with
\vth~ than for those without \vth.  The larger differences are for halo D, which 
after introducing random velocities to the particles produce a relatively hot core.  

Finally, from the measured density and dispersion velocity profiles, we calculate
the coarse--grained phase--space density profiles, $Q(r)$.  As seen in the lower
panels of  Fig. \ref{panel}, excepting  the innermost regions, the $Q(r)$ 
profiles are well described by a power law $Q \propto r^{-\alpha}$ with $\alpha
\sim -1.9$ close to that obtained for CDM halos by \citet{TN01}.  For the inner 
regions, the  $Q(r)$ profile of the halos simulated without \vth~ tends to steepen, 
specially in halos \Dnv~ and \Cnv.
The opposite happens for the halos simulated with adding \vth, the inner $Q(r)$ profile
tends to be flatter as \vth~ is larger. 
 
\section{Discussion} 
   
\subsection{Robustness of the results} 

The halos studied here are on one hand among the first virialized 
structures to form in our simulations, and on the other  their assembling 
process started relatively late in the universe, between $z\approx 0.6$ and $1$ 
\citep[for discussions about the mass assembling process of halos with masses close
to the damping scale in the power spectrum see e.g., ][]{moore99-cc,acv2001,bode2001,knebe02,
busha07}.
Because of the late collapse, one might argue that the halos studied here are not relaxed 
and therefore is not suprising to have the  deviations from 
the NFW profile, as  reported in \S 4.1 for the experiments with \vth=0.  
We have followed the evolution 
of some of our halos for more than a Hubble time (scale factor $a>1$), 
finding a negligible evolution in the density profiles since $a=1$. For example, halo 
\Dnv~ was run until $a=1.4$ (18.4 Gyr). The density profile of the halo at this 
epoch is practically the same as at $a=1$ (13.7 Gyr, see Fig. 4).   
As discussed in previos studies (see the references above), the collapse of 
halos of scales close to the damping scale seems to be quasi--monolithic (though 
highly non spherical). Thus, in regions that remain relatively isolated, as in 
the case of the halos selected for our study, halos suffer low mass accretion, and 
their structures remain almost unaltered since the initial collapse.

\begin{figure}[htb!]
\epsscale{1.0}
\vspace{9.0cm}
\includegraphics{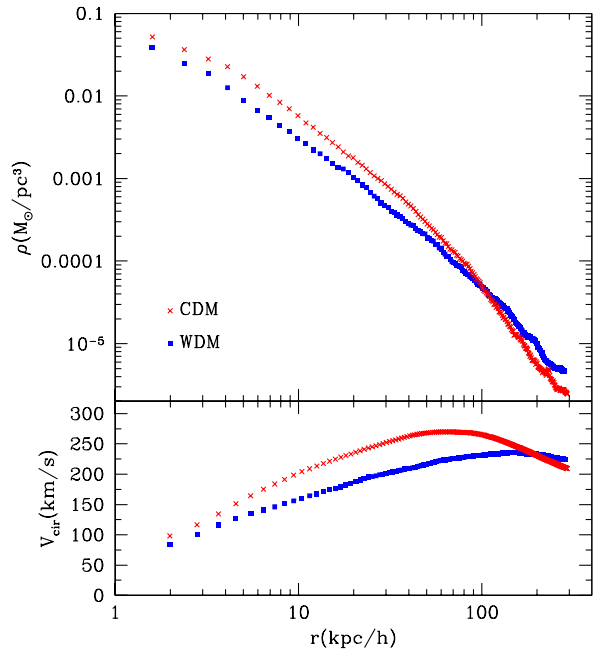}
\caption[fig:cdmwdm]{{\it Upper panel:} Comparison of the density profile
of a WDM Milky Way-like halo, without the random velocity, (squares)
with the density profile of its counterpart CDM, this latter generated
using the same random phases (crosses). {\it Lower panel:} Same as upper panel
but for circular velocity.
\label{cdmwdm}}
\end{figure}

We checked that the effects on the structure of halos reported in \S 4 are systematic 
by running for a WDM model --not shown in Table 1-- the corresponding CDM simulation,
using the same random phases and changing only the initial power spectrum.
We found that the density profile of the CDM halo 
is well fitted by the NFW function, while the corresponding WDM halo 
present the systematic deviation already seen in Fig. \ref{fig:perfnov}. 
Figure \ref{cdmwdm} compares the density and circular velocities profiles 
for the halo in question in its two versions, 
CDM and WDM without random velocities. Since the WDM halo 
ends up at $z=0$ with a slightly lower mass than the CDM halo
we correct the profiles so as to make the comparison at a fixed mass
($= 3.0 \times 10^{12}\ \msunh$). Notice, in particular, that the inner density 
profile of the WDM halo is indeed steeper than their CDM counterpart.

Concerning the resolution limit in our simulations,  based on the convergence study 
carried out for halo D (see Fig. 2), we find that a strong limit 
is $h_8=8 h_{\rm for}$, but resolution might be still acceptable for radii 
slightly larger than $h_4=4 h_{\rm for}$, a value suggested previously for CDM 
halos in ``equilibrium'' simulated with the ART code \citep{kkbp2001}.  With a 
resolution limit at $h_8$, our simulations allow to resolve the inner structure of halos 
down to $1.2$\kpch~ for the $512^3$ runs and down to $0.61$\kpch~ for the $1024^3$ run. 
These radii correspond respectively to $\sim 0.5$\% and 0.25\% virial radius in our halos.
 
Finally, it is important to recall that the masses of the the WDM halos 
analyzed in our simulations, are well above from the mass scale affected by 
discreteness effects, like the spurious formation of structures and 
substructures due to the initial grid pattern (see \S 3.1).

\subsection{Do soft cores form in WDM halos?}   

Early structure formation studies based on a WDM cosmology considered 
particles originated in thermal equilibrium. For this case, both \vth~ and \fsl~ depend only on 
the particle mass \mnu. The smaller \mnu, the larger \fsl~ and  \vth.  
Controlled numerical simulations of isolated halos showed that in order to produce 
``observable'' soft cores, the amplitude of \vth~ should be several times higher
than the values corresponding to thermal WDM particles of masses $\mnu\ga 1$ keV
\citep[][]{acv2001}. Particle masses smaller than $\sim 1$ keV are not allowed
by the constraints on satellite galaxy abundances as well as by the Ly--$\alpha$ 
power spectrum alone or  combined with CMBR and large scale structure data. 
The Ly--$\alpha$ power spectrum is the strongest of the constraints.  For
the non--thermal sterile neutrino, it places a limit on it mass at $\mnu\ga 2$ keV 
\citep{seljak2005,viel2006,kev2006b}. For thermal WDM particles, the observational 
constraints give a limit of $\mnu\ga 0.5$ keV \citep{narayanan2000,viel2005,kev2006b},
while a distinct analysis, using different simulations provide a stronger limit,
$\mnu\ga 2.5$ keV \citep{seljak2006}.

We may estimate the expected flat core radii of $2\times 10^{12}\msunh$ WDM halos
for thermal particles in the mass range $\mnu = (2-0.5)$ keV by using the approximation
given in \citet[][their eq. 13]{acv2001}. This approximation is based on the monolithic 
collapse of halos with non--negligible particle random velocities before the collapse.
For thermal WDM particles of
masses 2 and 0.5 keV, the $z_M$ of a $2\times 10^{12}\msunh$ perturbation are
$\approx 3.4$ and 1.9, respectively, while the $\vth(0)$ corresponding to these masses are 
0.1 and 0.015 \kms. Therefore, the expected core radii are $\rc\approx 30$ 
and 210 pc. 

The value of \vth~ for a non--thermal sterile neutrino of 0.5 keV is approximately three 
times larger than the corresponding to the thermal particle of the same mass. Therefore, 
for this case $\rc\approx 630$ pc. So, the resolutions that we may attain in our 
simulations of WDM halos for the
\mnu=0.5 keV sterile neutrino are already close to these estimates of the flat core radii.

Recently, alternative particle  models, like super--WIMPS, were proposed. 
The dark particles in these models may acquire random velocities non--thermally, for example, 
through the decay process of NLSP particles  \citep[e.g., charged Sleptons 
into gravitinos,][and see for more references Steffen 2006]{frt2003}. 
In these cases, \vth~ does not depend directly  on the damping scale of the 
linear power spectrum. However, an extra  parameter is introduced, the decaying epoch. 
Strategies to constraint this parameter using astrophysical observations have been proposed 
\citep{frt2003,strigari2007}. 

Some cosmological models  with super--WIMP particles may be in agreement with 
constraints 
based on structure formation, specially the Ly-$\alpha$ forest, and 
still allow for relatively large velocity dispersions, able to 
ameloriate the potential problems of \lcdm~ at small scales. This is the case of the so called
Meta--dark matter models that consider the late decay of neutralinos into gravitinos. These
models preserve a power spectrum similar to \lcdm~ models and at the same
time set a phase space limit in the innermost structure of dark halos 
due to the injection of random velocities to the 
particles  \citep{strigari2007, kaplinghat05}. However, as mentioned in the 
Introduction, it was still an open question of how efficient 
the introduction of \vth~  is  for producing significant effects on the
inner structure of simulated dark halos. One of the goals of the
present paper was just to expore this question by means 
of numerical simulations able to resolve the WDM halos down to $\sim 0.005$ \rvir.

The results presented in \S 4.2  show definitively that the addition of \vth~ 
flattens the inner density profiles of WDM halos, and more as \vth~ is higher. Differences 
in density profiles of halos simulated with $\vth > 0$ and those with $\vth = 0$,  
start to be evident at radii of 2--3 \kpch ($\sim 0.010\rvir-0.015\rvir$)  
for our $\approx 2-4\times 10^{12}\msunh$ halos.  The halo average inner density measured
at 0.01\rvir, $\bar{\rho_{1\%}}$ decreases on average by factors 1.5 and 2.5 for
the models with $\vth(0)=0.1$ and 0.3 \kms, respectively (column 11 in Table 1). Nevertheless, 
we notice that the concentration of the halos seems not to change significantly by the 
introduction of \vth.

In general, the NFW function does not describe
well the inner density profiles of our simulated halos; their profiles are much shallower
than $\gamma = -1$ as can be seen in Fig. \ref{profwithv}. 
However, is the size of the random velocity effect as theoretically expected?
The theoretical predictions are based on the existence of an upper limit 
in the fine--grained phase--space density due to the collisionless nature and 
finite relict velocity dispersion of particles.  This upper limit, $Q_{\rm 0,max}$, 
implies that the halo density profile must saturate and form a constant--density
core \citep{hd2000}.

For the random velocities used in this study ($\vth(0)=0.1$ and 0.3 \kms, corresponding
respectively to thermal and non--thermal 0.5 \kev~ sterile neutrinos), $Q_{\rm 0,max} = 3
\times 10^{-5}$ and $1.1 \times 10^{-6}$\ \funits. According to \citet{hd2000}, 
the core radius produced by the phase--space packing scales as $m_X^{-2} v_{c,\infty}^{-1/2}$, 
where $v_{c,\infty}$ is the asymptotic circular velocity for the assumed non--singular
isothermal sphere (their eq. 18). 
This implies that more massive halos have smaller, more tightly bound cores. 
Applying the same equation for $m_X = 0.5$ keV and $v_{c \infty} = 200\ \kms$, the 
core radius for the thermal particle  ($Q_{\rm 0,max} = 3\times 10^{-5}$\ \funits) is 85 pc, 
while for the  sterile neutrino ($Q=1.1 \times 10^{-6}$\ \funits) is 450 pc. 

In any case, our results can only marginally test these estimates. The resolution limit 
in our simulations with non--zero \vth~ is in between $\approx 1$ and 1.7 kpc.  If the inner 
density profile response to the introduction of \vth~ is gradual, one expect to see yet 
some effects at these radii, and this happen to be the case. 

A way to attempt to infer (extrapolate) the sizes of possible flat cores
in the simulated halos is by fitting the measured density profiles to an analytical 
function that implies a flat core.   Results of these fits were presented 
in \S 4.2 using the Zhao (1996) profile with $(\alpha,\beta,\gamma)=(0.7,3,0)$
as well as the one suggested by S2006. The latter function gives a poor description 
of the overall measured density profiles, which tend to be significantly less 
curved than the analytical model  (see Fig. \ref{profwithv}). 
The obtained (overestimated) values for \rc~ ($=0.1r_0$, see \S 4.2) are reported in 
column 13 of Table 3.  When the S2006 function is fitted to only the inner 30 kpc, 
the fits improve and the estimated core radii become smaller by 
roughly a factor of two (column 14). However, even for this case the core radii seems to 
be upper limits, with the exception of halo \Dv. 

We have found that the  WDM profiles are better described by the Zhao
function with $(\alpha,\beta,\gamma)=(0.7,3,0)$. The best fits to the measured
profiles give extrapolated core radii $\rc\approx 5-8$ times smaller than the S2006 profile
fitted to only the inner 30 kpc.  Should we have the sufficient resolution to
resolve the flat cores, their radii would lie in between $\rc$ and  $r_{c,30}^{S+}$.
The only halo for which the flat core is patent at our resolution limit is \Dv;
a visual inspection shows that the core radius is close to 1 \kpch.  

Our results show that the different estimates of the core radius increase by 
less than a factor of 1.5 from the simulations with \vth(0)=0.1 \kms~ to the ones 
with \vth(0)=0.3 \kms.  The amount of this increase is less than the one
we would predict using the monolithic collapse approximation of \citet{acv2001};
according to this approximation, the core radius of halos formed at the same
time depends linearly on the injected \vth~ at the maximum expansion of the perturbation,
$\rc\propto v_{\rm rms,z_M}= v_{\rm rms,0}(1+z_M)$.
We have estimated the predicted values of \rc~ for our halos by using this
approximation. From the simulations, we find that the redshifts of maximum
expansion of halos C, D, and E  are roughly $z_M = 1.6, 1.3$ and 1.8, respectively;
these redshifts are practically the same for the different values of \vth.
The calculated \rc~ for \vth(0) = 0.1 (0.3) \kms~ are then 221 (663), 240 (720), 
and 209 (627) pc, respectively. Thus, in general these predictions give
core radii just in between \rc~ and  $r_{c,30}^{S+}$ (see Table 1), though the
dependence on \vth~ is much more pronounced than for the (extrapolated) core radii 
estimated from the fits to our halos.

\subsection{How is the structure of halos at the damping scale?}
The simulations carried out in this paper allowed us also to explore the
structure of dark matter halos formed from perturbations at the scale of 
damping of the linear power spectrum. The cutoff in the power spectrum 
used here corresponds to a relatively large mass, $M_f=2.6\times 10^{12}$\msunh.  
Therefore, the formation of the first structures in this model, namely those 
structures with masses close to $M_f$, happens relatively late.
We speculate that the formation process of the truncation halos is generic. If this
is true, then the structure of the late--formed truncation halos simulated here with
several millions of particles (\vth=0) should be similar to the structure of early--formed 
truncation (micro)halos in models with much smaller filtering masses than the one
used here, for example in the CDM models. 
If this is the case, then our results may enrich the discussion about the formation 
and structure of the first microhalos in CDM models (Earth--mass scales).   

We have found a clear systematic trend in the density profiles of the simulated 
truncation halos:  they are significantly steeper than $r^{-1}$ in the inner 
regions, $r\lesssim 0.02\rvir-0.03\rvir$, and lie below the best NFW fits in the intermediate
region. CDM halos with an inner slope steeper than $r^{-1}$ have been reported 
in other contexts: recently merged group-- and cluster--size halos 
\citep{knebe02, tasitsiomi04} or microhalos formed at the scale of CDM 
power--spectrum damping  \citep{diemand2005}.
It has been argued that the recent major merger is the
one to blame for the steepening of the density 
profile while subsequent secondary infall modifies the external region.   
For the halos at the damping scale, rather than a major merger, 
the dynamical situation corresponds 
to a fast (cuasi--monolithic) collapse. However, in both cases, 
the process is dynamically violent.  We have checked that the 
obtained density profiles do not correspond to a transient 
configuration. As mentioned in \S 5.1, for halo D the profile remains almost 
unchanged until $a=1.4$.

Extrapolating our results to the damping scale of CDM, we can speculate that CDM 
microhalos may be significantly steeper than the NFW profile. This implies that
the possible contribution of surviving microhalos to the $\gamma-$ray flux originated 
by neutralino annihilation, might be comparable to the central flux from host halos \citep{dkm2006a}. 
Another implication of our results could be related to the buildup of the
inner density profile of dark halos in general. \citet{dehnen05} and \citet{kazantzidis06} 
argued that  the assembly of halo inner density profiles happens very early in the 
history of the universe and specifically that the inner slope of the
cuspier progenitor survives up to the final halo. If the density profiles 
in our simulated halos are representative of objects formed at the 
damping of power scales in general, then, according to the mentioned studies, the
central slope of present day dark matter halos should be much steeper than
the NFW profile.

In order to verify our results, a more systematic study  halo structure  at the
scale of damping is required, exploring any possible dependence with  the shape of the
cutoff and the power spectrum slope at the scale of damping. Currently the only 
studies discussing a similar situation report different results  for  the scale of
galaxy clusters\citep{moore1999b} and the microhalos \citep{diemandN05}. 
It is unclear if this suggests a slope dependence on the profile of the smallest 
dark matter halos with the power spectrum slope.

\section{Conclusions} 

We have studied by means of cosmological N--body simulations the structure
of dark halos formed in the context of a WDM model corresponding to 
a  non--thermal sterile neutrino particle of mass \mnu=0.5 keV.  The first series of 
simulations did not include the injection of a random velocity component, \vth, 
to the particles and were aimed at exploring the structure of the halos formed 
from perturbations at the 
damping scale in the linear power spectrum ($M_f=2.6 \times 10^{12}\ \msunh$
for the concrete WDM model studied here). The second series of simulations
included a \vth~ component  of a (i) a \mnu=0.5 keV thermal neutrino
($\vth(z=0) = 0.1$ \kms), and one that roughly corresponds 
(ii) to a \mnu=0.5 keV non--thermal sterile
neutrino ($\vth(z=0)\approx 0.3$ \kms). This latter model was
used to generate the initial power
spectrum.  These simulations were aimed at exploring the effect that a
random velocity component has on the inner structure of halos; in particular,
at dilucidating whether 
constant--density cores are produced or not in the simulations. 
The results of our study lead us to the following two main conclusions:

$\bullet$  The structure of halos formed from perturbations of scales close to $M_f$, and 
resolved with up to more than 16 millions of particles (with \vth=0),  is peculiar:  the inner 
density profile ($\la 0.02\rvir$) is systematically steeper than the best corresponding NFW fit 
(and the respective CDM counterpart), and the overall density profile ($> 0.02\rvir$) 
tends to be less curved than the best NFW fit; the outer profile slope is never steeper
than $-3$.  According to our tests, these differences with respect to the structure of halos 
assembled hierarchically, can hardly  be attributed to a peculiar dynamical state of the halos 
simulated here.

$\bullet$  The effect of adding \vth~ to the particles produces a significant flattening
of the inner density profile ($r\la  2-3 \kpch$ corresponding to $\sim 0.010\rvir-0.015\rvir$) 
of the simulated halos.  The different estimated (extrapolated) sizes of the nearly 
constant--density cores are of the order of the theoretical predictions, which give 
values below our resolution limit. For the halo masses simulated here,
$\mvir\approx (2-4)\times 10^{12}$ \msunh, the flat core radii estimated from
different fittings are between $\sim 0.1-0.8$ \kpch. An increase in \vth(0) from 
0.1 to 0.3 \kms~ produces an increase in the extrapolated core radii of a factor 1.5 or less. 
For one of our simulations (halo \Dv), the presence of a nearly constant--density core, 
of radius $\approx 1$ \kpch, is already revealed at the resolution limit; the same
halo simulated with a different random velocity seed is less flattened.

Although the simulations presented here refer to a concrete WDM model, they can be 
interpreted within a wide range of contexts.  The density profile of
dark halos with masses close to the truncation mass in the linear power spectrum is systematically
different from the NFW profile; in particular, the inner regions tend to be steeper.  These could
have important implications in the context of CDM models if a significant fraction of microhalos 
formed at the free--streaming CDM scales ($\sim 10^{-6}\msunh$) have survived until the
present epoch. In this case, the predicted $\gamma-$ray flux from the neutralino annihilation
in the center of these cuspy microhalos might be comparable to the central flux from host halos. 
On the other hand, the fact that the microhalos are so cuspy, could have some
interesting implications in the building up of the next hierarchies of the halo assembling,
as well as in the inner structure of the larger halos. 

Regarding the effects of random velocity injection to the particles, our results show 
evidence of significant inner flattening of the halo density profile at our resolution 
radii. These resolutions are not enough to test directly the predicted core radii 
by phase--space constraints \citep[e.g.,][]{hd2000} or by dynamical models of gravitational 
collapse with initial random velocities \citep[e.g.,][]{acv2001, bode2001}. However,
the inner extrapolations of the best--fit models to our simulated halos are consistent
with such predictions.

\acknowledgments
This work was supported by PAPIIT--UNAM grants IN112806-2 and IN107706-3
and by a bilateral CONACyT--DFG grant. The authors gratefully acknowledge the hospitality extended by 
the "Astrophysikalisches Institut Potsdam" where this paper was finished. OV acknowledges support from
the NSF  grant 02-05413  assigned to the UW during the initial stage of the project, and a CONACyT 
Repatriaci\'on fellowship.  Some of the simulations presented in this paper were performed using the 
HP CP 4000 cluster (Kan-Balam) at DGSCA--UNAM. We acknowledge the anonymous referee, whose helpful 
comments and suggestions improved some aspects of this paper.


\end{document}